\documentclass[hyper]{JHEP3}

\input{epsf}
\usepackage{epsfig}
\usepackage{amssymb}
\usepackage{amsfonts}
\usepackage{amsbsy}
\usepackage{amsmath,amssymb,amsthm,amscd,epsfig}
\usepackage[all]{xy}

\def\mod{{\rm mod}}
\def\p{\partial}
\def\pb{\bar{\partial}}

\def\IC{\mathbb{C}}

\def\IZ{{\mathbb{Z}}}
\def\IR{{\mathbb{R}}}
\def\IP{\mathbb{P}}
\def\IQ{\mathbb{Q}}

\def\CC {{\cal C}}
\def\CI {{\cal I}}
\def\CM {{\cal M}}

\def\CN {{\cal N}}

\def\CF {{\cal F}}

\def\CP {{\cal P }}

\def\CV {{\cal V}}

\def\CO {{\cal O}}

\def\CE {{\cal E}}

\def\CH {{\cal H}}

\def\half{\frac{1}{2}}

\def\ocm{{\overline{\cal M}}}

\def\one{{\hbox{ 1\kern-.8mm l}}}

\def\p{\partial}

\def\be{\bar{e}}

\def\half{\frac{1}{2}}
\def\be{\begin{equation}}
\def\ee{\end{equation}}
\def\ch{\mathrm{ch}}

\def\to{\longrightarrow}

\title{Crossing the Wall: Branes vs. Bundles}

\author{Emanuel Diaconescu and  Gregory~W.~Moore\\
 NHETC and Department of Physics and Astronomy,
Rutgers University,\\
Piscataway, NJ 08855--0849, USA\\
\\
{\tt duiliu@physics.rutgers.edu,  gmoore@physics.rutgers.edu} }

\abstract{ We test a recently proposed wall-crossing formula for the
change of the Hilbert space of BPS states in $d=4$, $\CN=2$
theories. We study decays of $D4D2D0$ systems into pairs of $D4D2D0$
systems and we show how the wall-crossing formula reproduces results
of G\"ottsche and Yoshioka on wall-crossing behavior of the moduli
of slope-stable holomorphic bundles over holomorphic surfaces. Our
comparison shows very clearly that the moduli space of the  D4D2D0
system on a rigid surface in a Calabi-Yau is \emph{not} the same as
the moduli space of torsion free sheaves, even when worldhseet
instantons are neglected.   Moreover, we argue that the physical
formula should make some new mathematical predictions for a future
theory of the moduli of stable objects in the derived category.  }

\begin{document}

\section{Introduction}

Consider a four-dimensional quantum field theory, or supergravity
theory, with $\CN=2$ supersymmetry on a spacetime which is
asymptotically Minkowskian. These theories have moduli
characterizing their vacua as well as distinguished subspaces in
their Hilbert space - spaces of BPS states - defined to be the
one-particle states transforming in small representations of the
supersymmetry algebra. One of the reasons the spaces of BPS states
are so useful and interesting is that the rigidity of the
representation theory of supersymmetry implies they are - like an
index - immune to many deformations of parameters. Nevertheless, in
$d=4, \CN=2$ theories closer inspection reveals that the  space of
BPS states   is only \emph{locally} constant, and in fact it depends
on the moduli of the vacuum, a feature which already played an
important role in Seiberg-Witten theory \cite{Seiberg:1994rs}.
Spaces of BPS states can jump discontinuously across real
codimension one walls in moduli space, known as walls of marginal
stability.

Recently, in the context of Calabi-Yau compactification of type II
string theory, a precise wall-crossing formula for the change in the
number of BPS states has been proposed \cite{Denef:2007vg}. It is
the purpose of this note to test that  formula in situations where
it is not obvious that the derivation of \cite{Denef:2007vg}
applies. Conversely, using the relation between BPS states and the
mathematics of coherent sheaves and their derived categories, we can
use the physically-derived formula to make some interesting
predictions for mathematics.

Here is a brief outline of the paper:  In section two we recall the
wall-crossing formula and suggest that it is a universal formula for
d=4 $\CN=2$ theories. In section three we apply it to the case of
wall crossing for type II strings on a Calabi-Yau manifold,
emphasizing the case where a $D4$ brane wrapping a surface $S$
splits as a pair of $D4$ branes wrapping $S$.  In section four we
turn to  the relation of D-branes to mathematical moduli spaces.  We
 review mathematical results on walls of stability for coherent
sheaves on surfaces. In section five we compare the physical formula
with the results of G\"ottsche and Yoshioka on wall-crossing
formulae for the Hodge polynomials of moduli spaces of coherent
sheaves on $S$, in the case where $S$ is rigid. The agreement turns
out to be perfect in the leading approximation as the K\"ahler class
goes to infinity. A surprising point emerges that - even neglecting
worldsheet instanton corrections - subleading corrections in the
expansion in large K\"ahler class lead to a distinction between the
physical and mathematical walls of stability. We interpret this as a
signal that the moduli space of D4D2D0 branes wrapping a rigid
surface $S$ is \emph{not} that of coherent sheaves - as is often
asserted - but rather that of stable objects in the derived
category. In section 6 we explore some generalizations which are of
interest both physically and mathematically. In particular in
section 6.1 we discuss decays of D4D2D0 systems into D4D2D0 systems
wrapping different surfaces. A surprising consequence of these
decays is that a D4D2D0 system can wrap an ample divisor and split
into two systems wrapping ample divisors, even at large K\"ahler
structure.\footnote{ Examples where an ample $D4$ decays  into a
pair of ample $D4$'s have been independently discovered in
\cite{DenefVandenBleeken}.} We comment on the implications of this
for the OSV conjecture in section 7, and conclude by pointing out an
interesting open problem.

\section{The Wall-Crossing Formula}

Let us recall the basic wall-crossing formula of ref.
\cite{Denef:2007vg} (whose notation and conventions we always
adopt).  First, we assume that the BPS state is a particle in a
spacetime which is asymptotically Minkowskian. We assume there is
some unbroken abelian gauge symmetry at low energy so that BPS
particles can be characterized by their electric and magnetic
charge. This charge, which we denote by  $\Gamma$, will be valued in
a symplectic lattice.  The moduli of the vacua will be denoted by
$t$, so we are interested in studying the spaces $\CH(\Gamma;t)$:
These are the finite dimensional spaces of BPS 1-particle states of
charge $\Gamma$ with boundary conditions at infinity corresponding
to the vacuum $t$.

The space $\CH(\Gamma;t)$ is a representation of the rotation group
$\textrm{Spin}(3)$. Because of the supersymmetry the representation
is of the form
\begin{equation}
\CH(\Gamma;t) = \biggl( 2(0) \oplus (\half)\biggr) \otimes
\CH'(\Gamma;t).
\end{equation}
In general we will let   $(j)$ denote a representation of
$\textrm{Spin}(3)$ of half-integer spin $j$. We interpret the space
for $j=-1/2$ as the zero vector space.
%

Next, we must introduce the Dirac-Schwinger-Zwanziger
duality-invariant symplectic product on the charges, denoted
$\langle \Gamma_1,\Gamma_2 \rangle$. We also need the central charge
of the $\CN=2$ supersymmetry algebra in the charge sector $\Gamma$
with vacuum determined by $t$. We denote this complex number by
$Z(\Gamma;t)$.

 The basic mechanism
by which $\CH(\Gamma;t)$ changes was already explained in
\cite{Cecotti:1992qh,Cecotti:1992rm,Seiberg:1994rs}. There are real
codimension one walls of marginal stability, denoted,
$MS(\Gamma_1,\Gamma_2)$ with $\Gamma = \Gamma_1 + \Gamma_2$ across
which those BPS states, which are boundstates of other BPS states of
charges $\Gamma_{1,2}$, become unstable. As with non-Fredholm
perturbations in index theory, a state can ``move off to infinity''
in fieldspace and leave the Hilbert space. The walls of marginal
stability are therefore defined by
\begin{equation}
MS(\Gamma_1,\Gamma_2) = \{ t\vert Z(\Gamma_1;t) = \lambda
Z(\Gamma_2;t)\not=0 \qquad\textrm{for\ some}\quad \lambda \in \IR_+
\}
\end{equation}
Now, a basic stability criterion was derived in
\cite{Denef:2000nb,Denef:2000ar} in the context of $d=4,\CN=2$
supergravity: A boundstate which decays across a marginal stability
wall will be stable on the side:
\begin{equation}\label{denefstability}
\langle \Gamma_1, \Gamma_2\rangle {\rm Im} Z(\Gamma_1;t)
\overline{Z(\Gamma_2;t)} > 0.
\end{equation}
The wall crossing formula then states that as $t$ moves through the
wall at $t_{ms}\in MS(\Gamma_1,\Gamma_2)$ from the stable side
(\ref{denefstability}) to the unstable side the space of BPS states
loses a summand
\begin{equation}\label{hilbspace}
\Delta \CH_{BPS}' = (j_{12}) \otimes \CH'(\Gamma_1;t_{ms}) \otimes
\CH'(\Gamma_2;t_{ms})
\end{equation}
where the spin $j_{12}$ is given by
\begin{equation}\label{spinformula}
j_{12}=- \half +  \half \vert \langle \Gamma_1,\Gamma_2\rangle
\vert.
\end{equation}

In stating (\ref{hilbspace}) we assume that $\Gamma_1, \Gamma_2$ are
primitive, and that the point $t_{ms}$ on the wall is generic in the
sense that it is not on the intersection of walls of marginal
stability for $\Gamma_i$ themselves. \footnote{In
\cite{Denef:2007vg}\ a generalization for the case when one of
$\Gamma_i$ is not primitive was proposed. The generalization when
both $\Gamma_1$ and $\Gamma_2$ are not primitive is open and appears
to be   challenging.}

Although the formula (\ref{hilbspace}) was derived within the
specific context of multi-centered solutions of supergravity we
believe the wall-crossing formula is in fact universal within the
context of $d=4, \CN=2$ theories and does not depend on being able
to represent the boundstate as a classical supergravity solution. On
a wall of marginal stability a boundstate of two BPS constituents is
marginally bound so the constituents can be adiabatically separated
from each other. By locality, the statespace should be a product of
the space of states for each constituent times the statespace for
the common electromagnetic field. A standard computation in
classical electromagnetism shows that two dyons in $\IR^3$ of charge
$\Gamma_1,\Gamma_2$ carry angular momentum around their midpoint
given by
\begin{equation}
\vec J =  \half \langle \Gamma_1,\Gamma_2\rangle \frac{\vec x_1 -
\vec x_2}{\vert \vec x_1 - \vec x_2\vert}.
\end{equation}
The ``correction'' by $-1/2$ in (\ref{spinformula}) above is a
quantum effect and can be established, in the context of
multi-centered solutions of supergravity,  as discussed in
\cite{Denef:2002ru}. It would be desirable to have a more general
argument for this quantum correction.

The space of BPS states is not only a representation of
$\textrm{Spin}(3)$ but also of the $U(1)$ $R$-symmetry, where the
supercharges have quantum numbers
\begin{center}
\renewcommand{\arraystretch}{1.3}
\begin{tabular}{|c||c|c|c|c|c|c|}
\hline   & $J_3 $ & $R$ & $ J_3+R  $ &$  J_3-R$
\\
\hline $Q_+$     & $\half $ & $ +\half $ & $1$ & $0$
\\
\hline  $Q_-$ & $-\half $ & $\half $ & $0$ & $-1$
\\
\hline $\bar Q_{\dot -} $   & $\half $ &$-\half $ & $0$ & $+1$
\\
\hline $ \bar Q_{\dot +} $   & $-\half $ &$-\half $ & $-1$ & $0$
\\
\hline
\end{tabular}
\end{center}\label{tab:squarereflections}

A useful corollary of (\ref{hilbspace}) for our discussion below
follows if we define
\begin{equation}
\Omega(\Gamma;t;x,y) := {\rm Tr}_{\CH'(\Gamma;t)} (-x)^{J_3+R}
(-y)^{J_3-R}
\end{equation}
where $J_3$ is a generator of ${\rm spin}(3)$ and  $R$ is the $U(1)$
$R$-charge of the BPS states.

Now, suppose the modulus $t$ crosses a wall where a particle of
charge $\Gamma$ can decay into constituents of charges
$\Gamma_1,\Gamma_2$.  The analysis of \cite{Denef:2002ru}, section
4.2 shows that all the states can be taken to have zero $R$-charge.
\footnote{Under the $R$-symmetry $\theta^\alpha \to e^{i \xi}
\theta^\alpha$ we need to have $W_\alpha \to e^{-i \xi} W_\alpha$
and hence $\lambda_\alpha \to e^{-i \xi} \lambda_\alpha$. If we take
the vacuum $\vert 0 \rangle$ in Denef's equation (4.14) to have
$R$-charge $+1$ then all the states in the Coulomb multiplet have
$R$-charge $0$.} Then the wall-crossing formula (\ref{hilbspace})
implies:

\begin{equation}\label{genwallcrossing}
\begin{split}
\Omega(\Gamma;t_+; x,y)  -&
\Omega(\Gamma;t_-;x,y)  =\\
   (-1)^{ \langle \Gamma_1,\Gamma_2\rangle-1}  (xy)^{- \half(\langle
 \Gamma_1,\Gamma_2\rangle-1)}  \frac{1-(xy)^{ \langle
 \Gamma_1,\Gamma_2\rangle}}{1-xy}   & \Omega(\Gamma_1;t_{ms};x,y)
\Omega(\Gamma_2;t_{ms};x,y)
\end{split}
 \end{equation}
where $t_+$ is on the side ${\rm Im}(Z_1 \overline{Z_2})>0$ and
$t_-$ is on the side ${\rm Im}(Z_1 \overline{Z_2})<0$.

Suppose that  the states in $\CH'(\Gamma;t)$ admit a description as
cohomology classes on some   moduli space $\CM$, which we assume is
K\"ahler and smooth. \footnote{In the mathematical applications,
smoothness is not obvious. In such cases we might be forced to
restrict attention to more primitive numerical invariants, such as
the Euler character.}  A 4D supersymmetric sigma model with K\"ahler
target space $\CM$, reduces
 to a $(2,2)$ supersymmetric quantum mechanics in $0+1$ dimensions.
 Under the identification of wavefunctions in the quantum
mechanics with differential forms on $\CM$ we have
\begin{eqnarray}
Q_+ & \rightarrow & \p \\
Q_- & \rightarrow & \pb^\dagger\\
\bar Q_{\dot -} & \rightarrow & \pb \\
\bar Q_{\dot +} & \rightarrow & \p^\dagger
\end{eqnarray}

In this situation we can relate $\Omega$   to the Hodge polynomial
$e(\CM;x,y)$ of $\CM$. We identify $J_3$ with the Lefshetz $sl(2)$
acting on the cohomology:
\begin{equation}
J_3 \omega = \half ( \deg \omega - \dim_{c} \CM)\omega
\end{equation}
and hence
\begin{eqnarray}\label{Hpolynomial}
\Omega(\Gamma;t;x,y)  &= & (-1)^{\dim\CM} (xy)^{-\half\dim \CM}
\sum_{p,q} (-1)^{p+q} x^{p} y^{q}\dim H^{p,q}(\CM)\\
&=& (-1)^{\dim\CM} (xy)^{-\half \dim \CM} e(\CM;x,y)
\end{eqnarray}
Evidently, for this equation to make sense, the moduli space $\CM$
must depend on $t$.

Two special cases are of particular interest: If we put $x=y$ then
we obtain the Poincar\'e polynomial. If we further take the limit
$y\to 1$ we obtain the Witten index, i.e., the Euler character of
$\CM$.

\section{Wall Crossing for Calabi-Yau Compactification of type II
strings}

Now let $X$ be a compact Calabi-Yau manifold, and consider the
compactification of type IIA strings on $X$. BPS charges are
elements of $\gamma\in K^0(X)$, but in this paper we will identify
the charge with its image in $H^{\rm even}(X;\IQ)$,
\begin{equation}
\Gamma = \ch(\gamma) \sqrt{Td(X)} := r + \ch_1(\gamma) + \hat
\ch_2(\gamma) + \hat \ch_3(\gamma).
\end{equation}
In this case, the symplectic product on charges is given by
\begin{equation}\label{symplectic}
\langle \Gamma_1,\Gamma_2\rangle = \int_X \Gamma_1 \Gamma_2^*
\end{equation}
where $\Gamma\to \Gamma^*$ reverses the sign of the components of
degree $2\mod 4$.  The relevant moduli space of vacua for IIA
strings is the complexified K\"ahler moduli space, and we identify
$t=B+iJ$ where $B$ is the flat $B$-field potential, $B\in
H^2(X;\IR)$ and $J\in H^2(X;\IR)$ is the K\"ahler class. In this
paper we will work in the limit of large K\"ahler class and ignore
worldsheet instanton corrections to the period vector. Thus we will
identify the holomorphic central charge with
\begin{equation}
Z_h(\Gamma;t)= - \int_X e^{-t} \Gamma.
\end{equation}
(We only use the central charge to compute walls of marginal
stability. Therefore, it suffices to use the holomorphic rather than
the normalized central charge. We henceforth drop the subscript
$h$.)

In order to compare with mathematical work we will, until section
\ref{sec:Generalizations}, concentrate on the case of D-branes which
are boundstates of $D4D2D0$ branes localized on a holomorphic
surface $S$ in $X$. As we review in section \ref{sec:Generalities}
below these are - classically- the pushforward from   $S$   of
coherent sheaves $E$ on $S$, or on a ``thickening of $S$.'' We will
furthermore take the sheaves on $S$ to be torsion-free.

In this case the charge is \cite{Green:1996dd,Minasian:1997mm}
\begin{equation}\label{charger}
\Gamma = \ch(j_*(E)) \sqrt{Td(X)}
\end{equation}
where $j: S \hookrightarrow X$ is the inclusion.  Let $c_1,c_2$ be
the Chern classes of $E$ and let $r$ be the rank. It is useful to
define
\begin{equation}\label{muanddelta}
\mu:=\frac{c_1}{r}, \qquad \Delta :=\frac{1}{r}\biggl(c_2 -
\frac{r-1}{2r} c_1^2\biggr)
\end{equation}
in terms of which
\begin{equation}\label{newcharge}
\Gamma = r [S] + r j_*\bigl( \hat \mu  \bigr) + q_0 \omega
\end{equation}
where $\hat \mu = \mu + \half c_1(S)$  and $\omega$ is the unit
volume form on $X$. The D0 charge is given by
\begin{equation}
q_0 = r\biggl[ \frac{\chi(S)}{24} + \int_S \half \hat \mu^2 - \Delta
\biggr]
\end{equation}

We now consider the wall of marginal stability for a decay $\Gamma
\to \Gamma_1 + \Gamma_2$ where all three charges
$\Gamma,\Gamma_1,\Gamma_2$ have the form (\ref{newcharge}). For
example $\Gamma, \Gamma_1,\Gamma_2$ could be the charges
corresponding to torsion-free sheaves $E,E_1,E_2$ on $S$, but there
are other possibilities, discussed in section \ref{sec:Generalities}
below.  In this case the wall of marginal stability can be computed
from the vanishing locus of \footnote{The solution set ${\rm Im}(Z_1
\bar Z_2)=0$ consists of both marginal stability and anti-marginal
stability walls, the latter being the case where the complex numbers
$Z_1$ and $Z_2$ anti-align. The asymptotic component of the wall
that we study is a marginal stability wall. }

\begin{eqnarray}\label{MSwallSingleSurf}
{\rm Im}(Z_1\overline{Z_2}) & = & \frac{r_1 r_2}{2} J_S^2 J_S\cdot
(\mu_1-\mu_2) + q_0^1 J_S \cdot r_2\hat \mu_2 - q_0^2 J_S\cdot
r_1\hat \mu_1 + \\
+& (J_S\cdot B_S) &  ( \frac{r_1-r_2}{2} J_S^2 + q_0^2-q_0^1) + r_1
r_2(J_S\cdot \hat \mu_1 B_S\cdot \hat \mu_2 - J_S \cdot \hat \mu_2
B_S \cdot \hat \mu_1) \\
& + & \frac{r_1 r_2}{2} B_S^2 J_S \cdot (\mu_2-\mu_1) + (J_S \cdot
B_S)(B_S\cdot(r_1\hat\mu_1 - r_2 \hat \mu_2))\\
& + & \frac{r_2-r_1}{2} B_S^2 (J_S \cdot B_S)
\end{eqnarray}
where we have organized terms so that each line is homogeneous in
$B$ and within each line we have written the highest order in $J$
first. In this formula all the intersection products are computed on
$S$. In particular,  $J_S, B_S$ denote the pullbacks of $J,B$ to the
surface $S$.

The spin factor is computed from\footnote{It is important to get the
sign right in this formula. Note that for a surface $S\subset X$ and
closed differential forms $\eta$, $\omega$ defined on $S$,$X$,
respectively we have $\int_X \omega \wedge j_*(\eta) = \int_S
j^*(\omega)\wedge \eta$. Now, in particular, if $PD(S)$ is the
Poincar\'e dual of $S$ then we may represent it as the Thom class of
the oriented normal bundle. Then, $j^*(PD(S))$ is the Euler class of
the normal bundle. For the case of a complex codimension one surface
the Euler class is $c_1(N(S\subset X))$. Now, because $X$ is
Calabi-Yau, $c_1(N(S\subset X)) = - c_1(TS)$. So $j^*(PD(S))=K_S$.}
\begin{equation}
\langle \Gamma_1,\Gamma_2\rangle = r_1 r_2 K_S \cdot (\mu_2-\mu_1)
\end{equation}
where $K_S$ is the canonical bundle of $S$.

The formulae simplify considerably if we restrict attention to the
subspace with $B=0$ and take $J_S \to \infty$. In this case the
marginal stability wall within the K\"ahler cone of $S$ can be
written as
\begin{equation}\label{curvewall}
J_S\cdot (\mu_1-\mu_2) = 2\biggl( \frac{q_0^2}{r_2} \frac{J_S\cdot
\hat\mu_1}{J_S^2} - \frac{q_0^1}{r_1} \frac{J_S\cdot
\hat\mu_2}{J_S^2}\biggr)
\end{equation}
which clearly asymptotes to the wall
\begin{equation}\label{straightwall}
J_S\cdot (\mu_2-\mu_1)=0
\end{equation}
for large $J_S$.

The distinction between eq. (\ref{curvewall}) and eq.
(\ref{straightwall}) is important, and is an order $1/J$ correction,
hence can be significant even when instanton corrections can be
neglected. This is a simple way of seeing that it is not sufficient
to use the category of coherent sheaves when describing
supersymmetric D-branes, and presumably the correct generalization
is to the derived category of coherent sheaves. We will return to
this point in section \ref{sec:CompareG}.

As a simple example, consider the case of $S=  \IP^1\times \IP^1$ so
that $J_S = x d_1 + y d_2$, with $d_i$ Poincar\'e dual to each
$\IP^1$ factor and hence $x,y>0$ in the K\"ahler cone. We have
plotted an interesting example in Figure 1.

\EPSFIGURE{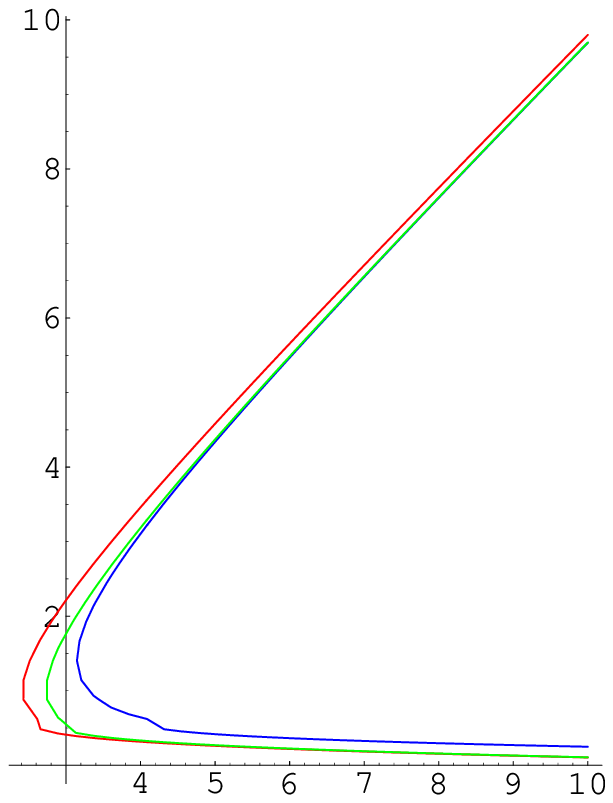,height=5cm,angle=0,trim=0 0 0 0}%
{Example of the walls at $B=0$ for different values of D0 charges
$q_0^1,q_0^2$ with the same total D0 charge. }
  \label{SplitGottscheWalls}

For large $J_S$,  Denef's stability conditions says that
\begin{equation}\label{DSspecial}
K_S\cdot (\mu_2-\mu_1) J_S\cdot (\mu_2-\mu_1)>0
\end{equation}
so this is the side on which there are ``extra'' states which we
will lose.

For rigid surfaces, the Poincar\'e dual $[S]$ will not lie in the
K\"ahler cone, so our discussion here goes beyond that of
\cite{Denef:2007vg}. (It is possible, but not obvious,  that
multicentered solutions corresponding to the above splits exist.) We
also remark that in equation (3.62) of version 1 of
\cite{Denef:2007vg}, the reader is led to the impression that D4
cannot split into D4+D4, but the argument there assumes that $[S]$
and $J_S$ are proportional to each other.

\section{Moduli Spaces }

\subsection{Generalities}\label{sec:Generalities}

The classical picture of a supersymmetric type IIA D-brane wrapping
cycles in a Calabi-Yau manifold  leads one naturally to the
identification of classical D-brane states with coherent sheaves on
$X$ \cite{Morrison:1996ac,Harvey:1996gc}. However, there is in fact
much evidence to suggest that the proper mathematical description of
the moduli of classical supersymmetric IIA branes on $X$ is given in
terms of the derived category of coherent sheaves \cite{ClayBook}.
Moduli spaces of derived objects have been constructed so far in
certain cases \cite{Inaba,TodaB} using the stability conditions of
Bridgeland \cite{BridgelandA}. Unfortunately, wall crossing formulas
seem to be out of reach at the present stage. Nevertheless, we can
make progress for the case of a \emph{rigid} holomorphic surface
$S\subset X$ since then the moduli space of $D4D2D0$ branes is
expected to be related to the moduli space of slope stable coherent
sheaves. However, as we will see below, even at large $J_S$, this is
not quite true.

In any case, one  should define a moduli space of stable objects of
fixed characteristic classes, corresponding to the charge $\Gamma$
and satisfying some $t$-dependent stability condition,
$\CM(\Gamma;t)$, and then, ideally, identify
\begin{equation}
\CH'(\Gamma;t) \sim H^*(\CM(\Gamma;t)).
\end{equation}
This equation is very rough. In addition to the actual construction
of the moduli space, one should specify what kind of cohomology one
is using since $\CM$ will in general be singular or noncompact, etc.

We can also turn things around, and use the physical formula to make
a prediction for wall-crossing behavior of the eventually-to-be
constructed moduli spaces of stable objects in the derived category.
For example,   even if $S$ is not rigid, we expect the moduli space
should be fibered over the moduli of holomorphic surfaces $S\subset
X$ with fiber given by the moduli of coherent sheaves on $S$ (again,
asymptotically for $J\to \infty$). In this way we make physical
predictions for more general moduli spaces of sheaves.

One subtlety we have thus far suppressed is the following. If we
consider a charge of the form (\ref{newcharge}) then when  $r>1$ the
$D4$-brane charge is not primitive and could in principle be the
charge of a ``thickening'' of the surface $S$. (By a ``thickening''
we mean that if $S$ is defined locally by the equation $f=0$ then
$rS$ is defined by the equation $f^r=0$.) Sheaves on such
thickenings can, and sometimes do, contribute extra components to
the moduli space. Physically, this corresponds to solutions to the
BPS embedding equations for the D-brane gauge field $A$ and normal
bundle scalars $\Phi$ in which $\Phi$ is nonzero. Such components
have been discussed, for example, in \cite{Donagi:2003hh}. In our
main application in section \ref{sec:CompareG} below $S$ will be
Fano, and using the vanishing theorem of \cite{Vafa:1994tf} (or an
analogous statement in the algebro-geometric version, due to R.
Thomas) one can show that such components do not occur.  Components
due to thickenings might be important in the more general
applications discussed in section \ref{sec:Generalizations} below.
Exploring this point should be interesting, but it is beyond the
scope of this paper. \footnote{We thank Richard Thomas for raising
the issue of these thickened components of moduli space. }

\subsection{Wall Crossing for moduli spaces of
coherent sheaves on a surface $S$. }
\label{sec:ModuliSpaces}

Let $S$ be a smooth projective surface with $-K_S$ effective, and
let $J_S$ be in the K\"ahler cone of $S$. Let
$\overline{\CM}(r,c_1,c_2;J_S)$ be the moduli space of rank $r\geq
1$ $J_S$-semistable torsion-free sheaves on $S$ with Chern classes
$(c_1,c_2)\in H^2(S, \IQ) \times H^4(X,\IQ)$. Note that torsion-free
sheaves of rank 1 are of the form $\CI_Z\otimes L$ where $Z$ is a
zero dimensional subscheme of $S$ and $L$ is a line bundle on $S$.
Such objects are stable for any polarization $J_S$, hence they do
not exhibit interesting wall crossing behavior. \footnote{It is
important that we restrict attention to $B=0$ here.}  Therefore we
will consider $r\geq 2$ from now on. We will denote by
$\CM(r,c_1,c_2;J_S)\subset \overline{\CM}(r,c_1,c_2;J_S) $ the open
subset corresponding to $J_S$-stable sheaves.

According to a theorem of Maruyama \cite{maruyama}, if $-K_S$ is effective,
$\CM(r,c_1,c_2;J_S)$ is smooth of expected dimension
\be\label{eq:expdimA}
\mathrm{dim}(\CM(r,c_1,c_2;J_S))=
2r^2\Delta -r^2\chi(\CO_S) +1
\ee
where $\Delta$ has been defined in (\ref{muanddelta}).

Therefore if any semistable sheaf $E$ with invariants $(r,c_1,c_2)$
is automatically stable, it follows that ${\overline
\CM}(r,c_1,c_2,J_S)$ is smooth and projective. This will be the case
if for example the rank $r$ and the degree $(c_1\cdot J_S)$ are coprime
(assuming that the polarization $J_S$ is integral). In such cases
the Hodge polynomial of ${\overline \CM}(r,c_1,c_2,J_S)$ is defined
as usual in terms of Dolbeault cohomology.

If there there exist strictly semistable objects $E$ with invariants
$(r,c_1,c_2)$, ${\overline \CM}(r,c_1,c_2;J_S)$ will be in general
singular. However, it turns out that ${\overline
\CM}(r,c_1,c_2;J_S)$ is by construction the GIT quotient of a closed
subscheme ${\overline Q}(r,c_1,c_2)$ of the appropriate $Quot$
scheme by an algebraic group $G=GL(N)$. According to
\cite{YoshiokaA,YoshiokaB}, if $-K_S$ is effective, the semistable
subset $Q(r,c_1,c_2)\subset {\overline Q}$ with respect to the
linearized group action is smooth. Then one can define a Hodge
polynomial of ${\overline \CM}(r,c_1,c_2;J_S)$ as the equivariant
Hodge polynomial of $Q(r,c_1,c_2)$ with respect to the $G$-action.
More precisely we have \be\label{eq:HodgepolA}
e(\CM_H(r,c_1,c_2),x,y) = {P_G(Q(r,c_1,c_2),x,y) \over 1-xy} \ee
where $P_G(Q(r,c_1,c_2),x,y)$ denotes the equivariant Hodge
polynomial of $Q(r,c_1,c_2)$. The normalization factor $1/(1-xy)$
represents the Hodge polynomial of the classifying space
$B\IC^\times$.

A more powerful approach has been recently developed in
\cite{FG,GNMY} for moduli spaces equipped with a perfect
tangent-obstruction complex. This allows one to define a virtual
$\chi_y$ genus as well as a virtual elliptic genus. It would be very
interesting to understand the physical applications of this
construction, but  we leave this for future work.

In the following we will adopt the definition of \cite{YoshiokaA,YoshiokaB}
assuming that $-K_S$ is effective, and show that the resulting
wall crossing is in agreement with physical predictions.

As shown for example in \cite{qin,Gottsche,YoshiokaB},
the moduli spaces of sheaves depend
on the choice of the polarization
$J_S$ in the K\"ahler cone and change discontinuously
across walls.
The dimension does not jump across walls except in cases when
the moduli space is empty on one side of a wall. Rather, the
two moduli spaces are related by birational transformations.

We will briefly recall some basics of wall crossing behavior
following the treatment of \cite{YoshiokaB}, which applies to higher
rank sheaves. Employing the notation of \cite{YoshiokaB}, let us
denote by $\gamma=(r,\mu,\Delta)$, where $\mu,\Delta$ has been
defined in (\ref{muanddelta}). The marginal stability walls in the
K\"ahler cone $\CC(S)$ are in one-to-one correspondence to sequences
$\gamma_i=(r_i,\mu_i, \Delta_i)\in H^0(S,\IQ)\oplus H^2(S,\IQ)
\oplus H^4(S,\IQ)$, $i=1,\ldots, s$, $s\geq 2$, satisfying the
following conditions
\begin{itemize}
\item[$(i)$]
There exists a filtration
\[0\subset \CF_1 \subset \cdots \subset \CF_s =E
\]
with $\gamma(\CF_i/\CF_{i-1})=\gamma_i$, $i=1,\ldots, s$.
\item[$(ii)$] There exists $H\in C(X)$ so that $(\mu_i-\mu_{i-1},H) =0$
for all $i=1,\ldots,s$.
\item[$(iii)$] $\Delta_i\geq 0$.
\end{itemize}
Given such a sequence $(\gamma_1,\ldots, \gamma_s)$, the wall $W$ is
defined by
\[
W = \{H\in C(X)| (\mu_i-\mu_{i-1},H) =0,\ i=1,\ldots, s \}.
\]
A chamber $\CC$ is defined to be a connected component of the
complement in $\CC(S)$ of the union of all walls $W$.
Since the moduli space does not vary within any given chamber,
we will write ${\overline \CM}(\gamma,\CC)$ for
${\overline \CM}(r,c_1,c_2,J_S)$ with $J_S\in \CC$.

Suppose $\CC_1,\CC_2$ are two chambers in the K\"ahler cone
separated by a wall $W$. Then according to \cite{YoshiokaB}, there
exist closed subsets $\CV_{\CC_a}\in {\overline \CM}(\gamma,
\CC_a)$, $a=1,2$ so that any $[E_1] \in V_{\CC_1}$ is slope-unstable
with respect to any polarization $J_{S,2}\in \CC_2$ and conversely
any $[ E_2] \in V_{\CC_2}$ is slope-unstable with respect to any
polarization in $J_{S,1}\in \CC_1$. Moreover, let
\[
0\subset HN_1(E_a) \subset \cdots \subset HN_{h_a}(E_a) =E_a
\]
be the corresponding Harder-Narasimhan filtrations\footnote{Any
unstable torsion-free sheaf $E$ on a smooth polarized projective
variety $X$ admits a canonical filtration $0\subset HN_1(E) \subset
\cdots \cdots \subset HN_h(E) =E$ called the Harder-Narasimhan
filtration of $E$ \cite{geommod}. This filtration is inductively
constructed so that each succesive quotient $HN_k(E)/HN_{k-1}(E)$ is
semistable and moreover $HN_k(E)/HN_{k-1}(E)$ is the maximal
destabilizing subsheaf of $E/HN_{k-1}(E)$. From a physical point of
view, the Harder-Narasimhan filtration encodes the decay products of
the unstable D-brane configuration described by $E$.} of $E_a\in
V_{\CC_a}$ for $a=1,2$. Then both filtrations have the same length
$h_1=h_2=s$ and the successive quotients satisfy
\[
\gamma(HN_{i}(E_1)/HN_{i-1}(E_1)) = \gamma_i,\qquad \gamma(
HN_{i}(E_2)/HN_{i-1}(E_2)) = \gamma_{s+1-i}.
\]
for $i=1,\ldots,s$.

In the following we will restrict ourselves to walls $W$
corresponding to length $s=2$ filtrations, which is the generic
situation. For a wall $W$ separating two chambers
$\CC_1,\CC_2$, we define $\Gamma_W$ to be
\[
\Gamma_W = \left\{ (\gamma_1,\gamma_2) \in
(H^{\textrm{ev}}(S,\IQ))^{2} \bigg| \begin{array}{l}
(\mu_1-\mu_2,J_S)=0\ \textrm{for\ all}\ J_S \in W\\
(\mu_1-\mu_2,J_{S,2})>0\ \textrm{for\ all}\ J_{S,2}\in
\CC_2\end{array}\right\}
\]
where $H^{\textrm{ev}}(S,\IQ) = H^0(S,\IQ) \oplus H^2(S,\IQ) \oplus
H^4(S,\IQ)$.
Let also
\[
d_{\gamma_1,\gamma_2} = -r_1r_2(P(\mu_2-\mu_1) -\Delta_1-\Delta_2)
\]
where $ P(x) = x\cdot(x-K_S)/2 +\chi(\CO_S)$. Then the wall-crossing
formula of \cite{YoshiokaB} reads\footnote{The wall crossing formulas
of \cite{YoshiokaB} are actually written for Poincar\'e polynomials.
The generalization to Hodge polynomials is straightforward.}
\begin{equation}\label{eq:wallcrossingA}
\begin{aligned}
& e(\ocm(\gamma,\CC_2),x,y) - e(\ocm(\gamma,\CC_1),x,y) =\\ &
{1\over 1-xy} \sum_{(\gamma_1,\gamma_2)\in \Gamma_{W}}
\big((xy)^{d_{\gamma_2,\gamma_1}} e(\ocm(\gamma_1,\CC_2),x,y)
e(\ocm(\gamma_2,\CC_2),x,y)\\
& \qquad \qquad \qquad \qquad -(xy)^{d_{\gamma_1,\gamma_2}}
e(\ocm(\gamma_1,\CC_1),x,y)
e(\ocm(\gamma_2,\CC_1),x,y)\big)\\
\end{aligned}
\end{equation}

Now we specialize to the case of rank $r=2$ sheaves. In this case
(\ref{eq:wallcrossingA})  is in agreement with the wall crossing
formula of \cite{Gottsche}. In this case we have
\begin{equation}
\gamma_1=(1,F_1,n_1) ,\qquad \gamma_2=(1,F_2,n_2)
\end{equation}
with $n_1,n_2\in \IZ_{\geq 0}$ and $F_1,F_2$ divisor classes on $S$.
In addition $F_1 +F_2 = c_1$. The  corresponding  two-term
Harder-Narasimhan filtration is
\begin{equation}
\CF_1=\CI_{Z_1}(F_1),\qquad E/\CF_1 = \CI_{Z_2}(F_2)
\end{equation}
where $Z_1,Z_2$ are zero dimensional subschemes of length $n_1,n_2$
respectively. The moduli spaces of rank 1 sheaves are insensitive to
the chamber structure. We have
\[
\begin{aligned}
\ocm(\gamma_1,\CC_1)&=\ocm(\gamma_1,\CC_2) = \mathrm{Pic}(S) \times
{\rm Hilb}^{n_1}(S) \\
\ocm(\gamma_2,\CC_1)&=\ocm(\gamma_2,\CC_2) = \mathrm{Pic}(S) \times
{\rm Hilb}^{n_2}(S) \\
\end{aligned}
\]
Let $\xi= \mu_1-\mu_2=2F_1-c_1$ as in \cite{Gottsche}. Then a
straightforward computation yields
\[
\begin{aligned}
d_{\gamma_1,\gamma_2} & = n_1+n_2 -\half \xi\cdot(\xi+K_S) -\chi(\CO_S)\\
d_{\gamma_2,\gamma_1} & = n_1+n_2 -\half \xi\cdot(\xi-K_S)-\chi(\CO_S) \\
\end{aligned}
\]
and one can easily check that the above formula \eqref{eq:wallcrossingA}
is in agreement with Theorem 3.4 of \cite{Gottsche}.

In this case we also have a very explicit description of the
closed subspaces of the moduli space of rank two sheaves which
become unstable when crossing the wall
$W^\xi=\{J_S\subset \CC(S)|\xi\cdot J_S=0\}$.
According to \cite{Gottsche} all rank two sheaves which become unstable
when crossing the wall $W^\xi$ from $\xi\cdot J_S <0 $ to
$\xi\cdot J_S>0$ are
extensions of the form
\begin{equation}\label{E12}
0 \rightarrow \CI_{Z_1}(F_1) \rightarrow \CE_{12} \rightarrow
\CI_{Z_2}(F_2) \rightarrow 0.
\end{equation}
%
%
%
For fixed $Z_1,Z_2$, $F_1, F_2$ the isomorphism classes of
extensions of the form (\ref{E12}) are parameterized by the
projective space $\IP {\rm Ext}^1(\CI_{Z_2}(F_2),\CI_{Z_1}(F_1)) $
with\footnote{The computation here is that $\chi(\CI_2,\CI_1) =
\int_S \ch(\CI_2^v) \ch(\CI_1) Td(S)$ and $\dim \textrm{Ext}^0=\dim
\textrm{Ext}^2=0$. }
\begin{eqnarray}\label{extE12}
K_{12}=\dim {\rm Ext}^1(\CI_{Z_2}(F_2),\CI_{Z_1}(F_1)) & =&  - \half
\xi
(\xi + c_1(S)) + n_1 + n_2 - \chi(\CO_S)\\
& = & - \half \xi\cdot c_1(S) + c_2 - \frac{c_1^2 + \xi^2}{4}
-\chi(\CO_S)
\end{eqnarray}
The closed subspace of the moduli space which destabilizes when
we cross the wall is isomorphic to a
closed subvariety $\CV_{12}$ of a projective
bundle $\CP_{12}$
\begin{equation}\label{eq:projbundleA}
\xymatrix{
\IP \textrm{Ext}^1(\CI_{Z_2}(F_2),\CI_{Z_1}(F_1)) \ar[r] & \CP_{12} \ar[d]\\
& \textrm{Hilb}^{n_1}(S) \times \textrm{Hilb}^{n_2}(S) \times
\textrm{Pic}(S) \times \textrm{Pic}(S)\\}
\end{equation}
Note that not all extensions of the form (\ref{E12}) are stable for
$\xi\cdot J_S<0$ \cite[Prop. 2.5(3)]{Gottsche}. Therefore $\CV$ will
in general be a proper subvariety of the above projective bundle.

There is a similar closed subspace  $\CV_{21}$ of the moduli space
which destabilizes when we cross the wall in the opposite direction.
This will be isomorphic to a closed subvariety of a projective
bundle $\CP_{21}$ of the form \be\label{eq:projbundleB} \xymatrix{
\IP\textrm{Ext}^1(\CI_{Z_1}(F_1),\CI_{Z_2}(F_2)) \ar[r] &  \CP_{21} \ar[d]\\
& \textrm{Hilb}^{n_1}(S) \times \textrm{Hilb}^{n_2}(S) \times
\textrm{Pic}(S) \times \textrm{Pic}(S)\\} \ee Although $\CV_{12},
\CV_{21}$ have in general positive codimension in the projective
bundles \eqref{eq:projbundleA}, \eqref{eq:projbundleB} respectively,
the proof of \cite[Thm. 3.4]{Gottsche} shows that
\be\label{eq:dimjump} \mathrm{dim}(\CV_{12}) -\mathrm{dim}(\CV_{21})
= \mathrm{dim}(\CP_{12}) - \mathrm{dim}(\CP_{21}). \ee

\section{Comparison with the physical wall crossing formula}
\label{sec:CompareG}

In order to compare the physical wall crossing formula with the
mathematical results reviewed in the previous section, note that any
smooth projective surface with $-K_S$ effective can be embedded in a
smooth projective Calabi-Yau threefold. The most obvious examples
are smooth elliptic fibrations with section over $S$. If $-K_S$ is
effective we can explicitly construct smooth Calabi-Yau Weierstrass
models over $S$ which admit a canonical section. Other examples can
be obtained by resolving del Pezzo singularities in Calabi-Yau
threefolds.

We will compare the physical and mathematical wall crossing formulas
assuming that for each $\gamma_a$, $a=1,2$, the moduli spaces
$\ocm(\gamma_a,\CC_1)$, $\ocm(\gamma_a,\CC_2)$ are isomorphic to
each other. In other words, we will assume that the moduli spaces of
the decay products do not change as we cross the wall. This is
automatic for decays of rank two sheaves as discussed at the end of
the previous section. Then we can denote these moduli spaces simply
by $\ocm(\gamma_a)$, $a=1,2$ omitting the polarization subscript.
Under this assumption, formula \eqref{eq:wallcrossingA} becomes
\begin{equation}\label{eq:wallcrossingB}
\begin{aligned}
& e(\ocm(\gamma,\CC_2),x,y) - e(\ocm(\gamma,\CC_1),x,y) =\\ &
\sum_{(\gamma_1,\gamma_2)\in \Gamma_{W}}
{(xy)^{d_{\gamma_2,\gamma_1}} -(xy)^{d_{\gamma_1,\gamma_2}}
\over 1-xy} e(\ocm(\gamma_1),x,y) e(\ocm(\gamma_2),x,y)
\\
\end{aligned}
\end{equation}
Using formula \eqref{eq:expdimA} for the expected dimension
of the moduli space, a straightforward computation yields
\be\label{eq:expdimB}
\begin{aligned}
& \mathrm{dim}(\ocm_{H_a}(\gamma)) - \mathrm{dim}(\ocm(\gamma_1))
-\mathrm{dim}(\ocm(\gamma_2)) =\\
& 2d_{\gamma_2,\gamma_1} +r_1r_2(\mu_2-\mu_1, K_S)-1
= 2d_{\gamma_2,\gamma_1} +\langle \Gamma_1,\Gamma_2\rangle -1.\\
\end{aligned}
\ee
Note also that
\[
d_{\gamma_1,\gamma_2} - d_{\gamma_2,\gamma_1} =
r_1r_2(\mu_2-\mu_1,K_S) =\langle \Gamma_1,\Gamma_2\rangle.
\]
Then \eqref{eq:wallcrossingB} can be further rewritten as
\be\label{eq:wallcrossingC}
\begin{aligned}
& e(\ocm(\gamma,\CC_2),x,y) - e(\ocm(\gamma,\CC_1),x,y) =\\ &
 \sum_{(\gamma_1,\gamma_2)\in
\Gamma_{W}}
(xy)^{(\textrm{dim}(\ocm(\gamma)) - \textrm{dim}(\ocm(\gamma_1))
-\text{dim}(\ocm(\gamma_2))- \langle \Gamma_1,\Gamma_2\rangle +1)/2}\\
& \qquad \qquad {1-(xy)^{\langle \Gamma_1,\Gamma_2\rangle}\over 1-xy}
e(\ocm(\gamma_1),x,y) e(\ocm(\gamma_2),x,y)
\\
\end{aligned}
\ee

Let us now  use the physical formula (\ref{genwallcrossing}) to work
out the change of the Hodge polynomial of ${\ocm}(r,c_1,c_2;J_S)$
when $J_S$ crosses a wall in the K\"ahler cone. Of course
$\Delta\Omega(\Gamma\to \Gamma_1+\Gamma_2;t_{ms})$ in general
depends on $t_{ms}$. We are assuming this has a finite limit for
$J_S \to \infty$. From the spin factor we get:
\begin{equation}\label{spinfactor}
{\rm sign}(K_S\cdot(\mu_1-\mu_2))
(-1)^{\langle \Gamma_1,\Gamma_2\rangle +1}
(xy)^{-\half \langle \Gamma_1,\Gamma_2\rangle +\half}
\frac{1-(xy)^{\langle \Gamma_1,\Gamma_2\rangle}}{1-xy}.
\end{equation}
Equation (\ref{Hpolynomial}) yields
\[
e(\ocm(\gamma),x,y) = (-1)^{\textrm{dim}(\ocm(\gamma,J_S))}
(xy)^{\half \textrm{dim}(\ocm(\gamma,J_S))} \Omega(\Gamma;t,x,y)
\]
for any charge $\Gamma$. Therefore the physical wall crossing formula
(\ref{genwallcrossing}) yields the following prediction for
the change in the Hodge polynomials
\be\label{eq:wallcrossingD}
\begin{aligned}
\textrm{sgn}(K_S\cdot(\mu_1-\mu_2))
& (-1)^{\textrm{dim}(\ocm(\gamma)) - \textrm{dim}(\ocm(\gamma_1))
-\text{dim}(\ocm(\gamma_2))- \langle \Gamma_1,\Gamma_2\rangle +1}\\
& (xy)^{(\textrm{dim}(\ocm(\gamma)) - \textrm{dim}(\ocm(\gamma_1))
-\text{dim}(\ocm(\gamma_2))- \langle \Gamma_1,\Gamma_2\rangle +1)/2}\\
& {1-(xy)^{\langle \Gamma_1,\Gamma_2\rangle}\over 1-xy}
e(\ocm(\gamma_1),x,y) e(\ocm(\gamma_2),x,y)\\
\end{aligned}
\ee
The overall sign can be simplified using the second equation in
\eqref{eq:expdimB} which shows that the exponent of $(-1)$ in the first
line of \eqref{eq:wallcrossingD} equals $2d_{\gamma_1,\gamma_2}$.
Since $d_{\gamma_1,\gamma_2}$ is an integer by construction
(it is actually the codimension of a certain subscheme of a
$Quot$ scheme \cite{YoshiokaB}), we are left with
\be\label{eq:wallcrossingE}
\begin{aligned}
\textrm{sgn}(K_S\cdot (\mu_1-\mu_2))
&(xy)^{(\textrm{dim}(\ocm(\gamma)) - \textrm{dim}(\ocm(\gamma_1))
-\text{dim}(\ocm(\gamma_2))- \langle \Gamma_1,\Gamma_2\rangle +1)/2}\\
& {1-(xy)^{\langle \Gamma_1,\Gamma_2\rangle}\over 1-xy}
e(\ocm(\gamma_1),x,y) e(\ocm(\gamma_2),x,y)\\
\end{aligned}
\ee

Now let us compare formulas \eqref{eq:wallcrossingC} and
\eqref{eq:wallcrossingE}. In \eqref{eq:wallcrossingC} there is a sum
over all pairs of charges $(\gamma_1,\gamma_2)$ in the set
$\Gamma_W$ associated to a given wall $W$. This reflects the fact
that the moduli space undergoes simultaneous birational
transformations associated to all possible two term destabilizing
Harder-Narasimhan filtrations. On the other hand, note that the
left-hand-side of (\ref{curvewall}) is independent of the choice of
distribution D0 brane charge between the two decay products, which
is encoded in the invariants $\Delta_1,\Delta_2$. However, the
right-hand-side is \emph{not} independent, and the different torsion
free sheaves with the same total second Chern class will in fact
lead to different physical stability  walls.

In order to emphasize this point, let us concentrate
on moduli spaces of rank 2 sheaves as in \cite{Gottsche}.
Indeed, in this case, the right-hand-side of (\ref{curvewall})
  reads:
\begin{equation}
\frac{1}{J_S^2}(J_S\cdot(\mu_2-\mu_1)) (n_1+n_2)+ \frac{1}{J_S^2}
(J_S \cdot (c_1+c_1(S))) (n_1-n_2) + \cdots
\end{equation}
where $+\cdots$ is independent of $n_1$ and $n_2$.  The dependence
of the walls on $n_1-n_2$ at fixed $n_1+n_2$ is illustrated in
Figure 1.

\emph{Thus, already at large radius, taking into account the leading
correction in the $1/J$ expansion, but not including instanton
effects, one sees that the relevant physical moduli space cannot be
the moduli space of coherent sheaves!} We interpret this as a signal
that the physical moduli space should be   the ``moduli space of
stable objects in the derived category.'' As we have mentioned, such
a moduli space has not been constructed, so we can take the physical
formula as a prediction for what should be true about such moduli
spaces.

Let us finally  compare the sign of the wall-crossing formula. Using
Denef's stability condition (\ref{denefstability}), which becomes
(\ref{DSspecial}) in our case, we  see that we \emph{lose} the
factorized Hilbert space as we go \emph{from}
\[
(K_S\cdot(\mu_1-\mu_2))(J_S
\cdot (\mu_1-\mu_2)) <0\qquad  \mathrm{to}\qquad
 (K_S\cdot(\mu_1-\mu_2))(J_S\cdot (\mu_1-\mu_2))>0.\]
However the mathematical wall crossing formula \eqref{eq:wallcrossingC}
claims a universal result for $\Delta e$ going from
\[
J_S \cdot (\mu_1-\mu_2)
>0\qquad  \mathrm{to}
\qquad  J_S\cdot (\mu_1-\mu_2) <0.\]
These are in beautiful agreement, since the
spin factor indeed changes sign if we change
$K_S\cdot (\mu_1-\mu_2)<0$ to
$K_S\cdot (\mu_1-\mu_2)>0$.
To check the absolute sign note that   $K_S\cdot (\mu_1-\mu_2)<0$
 corresponds to going from $J\cdot (\mu_1-\mu_2)>0$
(this is the moduli space denoted $M_{\CC'}^\gamma$ in Corollary 3.3 of
\cite{YoshiokaB})  to  $J_S\cdot (\mu_1-\mu_2)<0$
(this is the moduli space denoted $M_{\CC}^\gamma$
in Corollary 3.3 of \cite{YoshiokaB}).
To compute the $e$-trace over the states we lose we
 therefore compute $e(M_{\CC'}^\gamma) - e(M_{\CC}^\gamma)$.
The agreement between
 the two formulae is perfect!

\section{Generalizations}\label{sec:Generalizations}

We have argued that the wall-crossing formula is universal, and
hence we expect the physical formulae to apply to a wide range of
situations which look very different from the mathematical point of
view. Here we just point out a few special cases where the
mathematical counterparts are unknown, but perhaps within reach.

\subsection{Bundles on different surfaces}\label{sec:D4toD4D4}

We first generalize the story to decays where $\Gamma\to \Gamma_1 +
\Gamma_2$ involves $D4$ splitting into a pair of $D4$'s, but now the
support of the two constituent $D4$'s are in \emph{different}
cohomology classes. Thus, we can no longer work within the framework
of holomorphic bundles on surfaces, but must consider torsion
sheaves within the Calabi-Yau $X$. The surfaces have  Poincar\'e
duals     denoted by $S_1,S_2$, respectively. Suppose the surfaces
wrap $\Sigma_1,\Sigma_2$ and let $j_i: \Sigma_i \to X$ be the
inclusion.

Now we have
\begin{equation}
{\rm Im}(Z_1\overline{Z_2})  = \frac{r_1 r_2}{2}\biggl[ J_{S_2}^2
J_{S_1}\cdot {\widehat \mu}_1 - J_{S_1}^2 J_{S_2}\cdot
{\widehat \mu}_2 \biggr] + \cdots
\end{equation}
\begin{equation}
\langle \Gamma_1, \Gamma_2\rangle = r_1 r_2 \bigl( j_2^*(S_1)\cdot
{\widehat \mu}_2 - j_1^*(S_2) \cdot {\widehat \mu_1}\bigr)
\end{equation}

Even to leading order in  $J$ the walls are now in general nonlinear
and given by
\begin{equation}\label{eq:walleqnA}
\frac{J_{S_1}\cdot \hat \mu_1}{J_{S_1}^2} =\frac{J_{S_2}\cdot \hat
\mu_2}{J_{S_2}^2}.
\end{equation}

There are many examples in this class because linear systems on
compact threefolds generically contain reducible divisors. For
concreteness we will consider here decays associated to
degenerations of spectral covers in an elliptic fibration $X$. Such
divisors are of special interest because they are related to
torsion-free sheaves supported on the Calabi-Yau threefold $X$ by
Fourier-Mukai transform. Therefore the physical wall crossing
predictions for spectral covers can be translated to similar
statements concerning bundles supported on the Calabi-Yau threefold
$X$.

Let $\pi:X\to B$ be a smooth elliptic fibration with a section over
a base $B$, where we take $B$ to be a smooth projective surface with
effective anticanonical class. We will denote by $\sigma$ the
section class on $X$ and write the K\"ahler form of $X$ in the form
\[
J = t_f\alpha_f + \pi^* J_B
\]
where $\alpha_f = \sigma + \pi^*c_1(B)$ is Poincar\'e dual to
the elliptic fiber class. Note that we have the following relations
in the intersection ring of $X$
\[
\alpha_f \cdot \sigma =0 \qquad \alpha_f^2 = \alpha_f \cdot \pi^*c_1(B)
\qquad \alpha_f\cdot\pi^*\omega_1\cdot \pi^*\omega_2 = (\omega_1\cdot
\omega_2)_B
\]
for any curve classes $\omega_1,\omega_2$ on $B$.

Consider the linear system $|m\sigma +\pi^*\eta|$ where
$\eta$ is an effective curve class on $B$ and $m\geq 2$.
The generic member in this linear system is smooth and irreducible
if $\eta$ satisfies the following conditions
\cite{DLOPI,DLOPII,DOPWI,DOPWII,OPP}
\begin{itemize}
\item[$(i)$] $|\eta|$ is a base-point free linear system on $B$
\item[$(ii)$] $\eta-mc_1(B)$ is an effective curve class on $B$
\end{itemize}
Both these conditions will be satisfied if $\eta$ is a sufficiently
ample curve class on $B$.
We will assume this to be the case from now on.
Note that this also implies that the generic divisor in the
above linear system is ample on $X$.
We will be interested in moduli spaces of torsion coherent sheaves on
$X$ supported on divisors in $|m\sigma+\pi^*\eta|$. These moduli
spaces contain closed subsets parameterizing isomorphism classes
of sheaves with reducible support. In the following we will
show that the sheaves with reducible support can become unstable
by crossing certain walls in the K\"ahler cone of $X$,
provided that certain numerical conditions for Chern classes are satisfied.

With this goal in mind let us consider a configuration of two smooth
irreducible divisors $\Sigma_1, \Sigma_2$ with classes
$S_1=m_1\sigma + \pi^*\eta_1$, $S_2=m_2\sigma + \pi^*\eta_2$ where
$m_1+m_2=m$, $\eta_1+\eta_2=\eta$. We will assume that $\Sigma_1,
\Sigma_2$ intersect transversely along a smooth curve in $X$.
According to \cite{reducible}, coherent sheaves supported on the
union $\Sigma_1\cup \Sigma_2$ are in one-to-one correspondence with
pairs $(E_1,E_2)$ of coherent sheaves supported on $\Sigma_1,\
\Sigma_2$ respectively and a morphism $f:E_1|_{\Sigma_1\cap
\Sigma_2} \to E_2|_{\Sigma_1\cap \Sigma_2}$. Each sheaf $E_i$ has
topological invariants $(r_i,\mu_i,\Delta_i)$ as in the previous
section for $i=1,2$. We will take $\mu_1,\mu_2$ to be some generic
 divisor classes on $S_1$ respectively $S_2$ obtained by
pull-back from the ambient space
\[
\mu_1= b_1\sigma +\pi^*\rho_1,\qquad
\mu_2= b_2\sigma + \pi^*\rho_2,
\]
where $b_1,b_2\in \IQ$.
Then we have
\be\label{eq:slopefctsA}
\begin{aligned}
J_{S_i}^2 = &  (\eta_i\cdot c_1(B))t_f^2 + 2(\eta_i\cdot
J_B)t_f + m_iJ_B^2 \\
J_{S_i}\cdot {\widehat \mu}_i  =&
((\eta_i\cdot\rho_i)- \half \eta_i^2)t_f+(b_i\eta_i -b_im_ic_1(B)+m_i\rho_i)\cdot J_B \\
&   +\half m_i^2 c_1(B)\cdot J_B - m_i\eta_i\cdot J_B \\
\end{aligned}
\ee
for $i=1,2$,
where all intersection numbers are computed on $B$.
Now the walls can be found by substituting equations
\eqref{eq:slopefctsA} in \eqref{eq:walleqnA}.
In general we will obtain a fairly complicated cubic equation
for the K\"ahler parameters.

The spin factor can be computed from
\begin{equation}
\langle \Gamma_1,\Gamma_2\rangle = r_1 r_2 S_1\cdot S_2 \cdot
\biggl[ ((b_2-\half m_2)- (b_1-\half m_1))\sigma +
\pi^*(\rho_2-\rho_1 - \half (\eta_2-\eta_1)) \biggr]
\end{equation}

In order to simplify the computations let us specialize the
discussion to the case $B={\IP}^2$. Let $h$ denote the hyperplane
class of $B$. Then we can write
\[
J_B =t_b h, \qquad \eta_i =n_ih,\qquad \rho_i=a_ih
\]
for some positive integers $n_i\in \IZ$ and $a_i\in \IQ$,
$i=1,2$. Equations \eqref{eq:slopefctsA} become
\be\label{eq:slopefctsB}
\begin{aligned}
J_{S_i}^2 & = m_it_b^2 + 3n_it_f^2 + 2n_it_bt_f\\
J_{S_i}\cdot {\widehat \mu}_i & = -\half n_i(n_i-a_i)t_f +
\left(n_ib_i+(a_i-3b_i-n_i)m_i+{3\over 2}m_i^2\right)t_b.\\
\end{aligned}
\ee
Substituting in \eqref{eq:walleqnA}, we obtain the following
cubic equation
\be\label{eq:walleqnB}
\begin{aligned}
& {{-\half n_1(n_1-a_1)x +
\left(n_1b_1+(a_1-3b_1-n_1)m_1+{3\over 2}m_1^2\right)}\over
3n_1x^2 + 2n_1x+m_1} =\\
&{{-\half n_2(n_2-a_2)x +
\left(n_2b_2+(a_2-3b_2-n_2)m_2+{3\over 2}m_2^2\right)}\over
3n_2x^2 + 2n_2x+m_2}
\end{aligned}
\ee
 where $x=t_f/t_b$.

Marginal stability walls will correspond to positive real solutions
of equation \eqref{eq:walleqnB}. This yields several conditions on
the Chern classes which can  in principle be satisfied because we
have many free parameters $(m_i,n_i,a_i,b_i)$, $i=1,2$. In order to
obtain a more tractable equation, let us make a further
simplification taking the component $S_1$ to be the section of the
elliptic fibration i.e. $m_1=1$, $n_1=0$. We will also set $b_1=0$.
Then $n_2=n$,   we obtain the quadratic equation
\be\label{eq:walleqnC} 3nx^2 + n\left(2+{n-a_2\over
2a_1+3}\right)x+m_2 -{2nb_2+2(a_2-3b_2-n)m_2+3m_2^2 \over 2a_1+3 }=0
\ee
 if $a_1\neq -3/2$, and the linear equation
\be\label{eq:walleqnD} \half n(n-a_2) x =
\left(nb_2+(a_2-3b_2-n)m_2+{3\over 2}m_2^2\right) \ee if $a_1=-3/2$.

Now it is clear that these equations will have positive real
solutions in a certain range of the parameters
$(n,m_2,a_1,a_2,b_2)$. For each such solution the corresponding wall
is a straight line in the K\"ahler cone.

An interesting special case is $m=1$. In this case all divisors
in a linear system of the form $|\sigma + \pi^*\eta|$ are reducible
if $\eta\neq 0$. The generic divisor has two components --
a horizontal component in class $S_1=\sigma$ and a vertical component
in class $S_2=\pi^*\eta$. Let us take
\[
\mu_2 = b_2\sigma + cf
\]
where $f$ is an elliptic fiber class of the vertical component
$\Sigma_2$. Note that this is not a generic divisor class obtained
by restriction from $X$ as in the previous example. We will keep
$b_1=0$. Repeating the previous computations we obtain
\be\label{eq:slopefctsC}
\begin{aligned}
J_{S_1}^2 & = J_B^2 \\
J_{S_2}^2 & = (\eta\cdot c_1(B))t_f^2 + 2(\eta\cdot J_B)t_f\\
J_{S_1}\cdot {\widehat \mu}_1 & = \mu_1\cdot J_B +\half c_1(B)\cdot J_B \\
J_{S_2}\cdot {\widehat \mu}_2 & =
-\half \eta^2t_f + b_2(\eta\cdot J_B) + ct_f\\
\end{aligned}
\ee
Specializing again to the case $B={\IP^2}$ we obtain the
quadratic equation
\be\label{eq:walleqnE}
3nx^2 +\left(2n+{n^2-2c\over 2a_1+3}\right)x -{2nb_2\over
2a_1+3} =0
\ee
assuming again $a_1\neq -3/2$. If $a_1=-3/2$ are left again with a
linear equation
\[
\left(n^2-2c\right)x = 2nb_2.
\]

To conclude this section let us briefly translate the above wall
crossing predictions into similar statements for torsion-free sheaves
on $X$ using the Fourier-Mukai transform
\cite{FMWA,FMWB,BJPS,RonA,RonB,AndreasA,AndreasB}.
According to \cite{FMWB,BJPS} reducible spectral
covers correspond to bundles constructed by extensions.

More precisely, suppose the spectral data consists of
two smooth irreducible divisors $\Sigma_1,\Sigma_2$ equipped
with spectral line bundles $L_1,L_2$, and an
isomorphism $L_1|_{\Sigma_1\cap \Sigma_2}
\simeq L_2|_{\Sigma_1\cap \Sigma_2}$
as in section 5 of \cite{BJPS}.
The intersection $(\Sigma_1\cap \Sigma_2)$ is assumed
transverse and smooth.
Let $F_1,F_2$ be the holomorphic
bundles on $X$ corresponding to the spectral
data $(\Sigma_1,L_1)$, $(\Sigma_2,L_2)$ respectively.
Then the corresponding bundle $F$ on $X$ is obtained by an elementary
modification of the form
\be\label{eq:FMextA}
0\to F \to F_1\oplus F_2 \to Q \to 0
\ee
where $Q$ is a torsion coherent sheaf on $X$ supported on the vertical
divisor $D=\pi^{-1}(\pi(\Sigma_1\cap \Sigma_2)))$.
$Q$ is essentially the Fourier-Mukai transform of the sheaf
$L_1|_{\Sigma_1\cap \Sigma_2}$.

The previous computations predict that bundles of the form
\eqref{eq:FMextA} will become unstable as we cross certain walls of
marginal stability whenever equation \eqref{eq:walleqnB} admits real
positive solutions. In certain cases, such elementary modifications
can be equivalently described as extensions. For example suppose
that $Q = j_{D*}(F_1|_D)$ where $j_D:D \hookrightarrow X$  is the
embedding of $D$ in $X$.  Then $F$ is isomorphic to an extension of
the form \be\label{eq:FMextB} 0\to F_1(-D) \to F \to F_2\to 0. \ee
In these cases, the extensions become unstable when crossing the
wall, yielding a higher dimensional analogue of the decays studied
in the previous sections.

It is very interesting to consider the case $m=1$ from this point of view.
In this case the spectral data consists of a
horizontal component $\Sigma_1$ identified with the canonical section,
and a vertical component $\Sigma_2$. We also have
line bundles which agree on the intersection as above.
Using equations \eqref{eq:slopefctsC}, \eqref{eq:walleqnC}
it is not hard to produce concrete examples of marginal
stability walls for such configurations.

As shown in section 5 of \cite{OPP}, the
Fourier-Mukai transform of this spectral data is
a rank one torsion free sheaf of the form $\CI_Z\otimes L$
where $Z$ is a codimension two subscheme of $X$ and $L$
is a line bundle on $X$.
Note that the ideal sheaf $\CI_Z\otimes L$ fits in an exact sequence
\[
0\to \CI_Z\otimes L \to L \to \CO_Z\otimes L \to 0.
\]
Then our prediction is that $\CI_Z\otimes L$ will become unstable
across the wall, and it will decay into $L$ and $\CO_Z\otimes L$.
This leads to an apparent contradiction since rank one torsion-free
sheaves are known to be stable for any values of the K\"ahler
moduli. Here we predict nontrivial wall-crossing behavior even for
trivial $B$-field, generalizing the examples found in
\cite{Denef:2007vg}. Although such decays are impossible in the
abelian category of coherent sheaves on $X$, they are very natural
from the point of view of $\Pi$-stability in the derived category of
$X$ \cite{DouglasA,DouglasB,AD}. Indeed, the Fourier-Mukai transform
is related to $T$-duality in the physical setup, and hence the
K\"ahler class of the fiber will not be large. Accordingly one
cannot neglect worldsheet instanton corrections and one must use
$\Pi$-stability. It would be interesting to study this in detail
using the rigorous methods developed in \cite{BridgelandA}. Similar
decays of ideal sheaves as well as applications to enumerative
geometry are being considered in \cite{PT}.

\subsection{Bundles on $X$: $D6 \to D6 + D6$}

Suppose $\CE \to X$ is a general torsion free sheaf  on $X$. We set
\begin{equation}
\Gamma = \ch(\CE) \sqrt{Td(X)} := r + \ch_1(\CE) + \hat \ch_2(\CE) +
\hat \ch_3(\CE).
\end{equation}
For decays of a $D6$ to a pair of $D6$ branes, the marginal
stability wall will be a subset of the vanishing locus of

\begin{eqnarray}\label{D6Wall}
{\rm Im}(Z_1\overline{Z_2}) & = & \frac{J^3}{12}\biggl( r_2 J^2
\ch_1(\CE_1) - r_1 J^2\ch_1(\CE_2)\biggr)
+ \frac{J^3}{6} (r_1\hat\ch_3(\CE_2) - r_2\hat\ch_3(\CE_1))\\
& + &
\half J^2\ch_1(\CE_2) J \hat \ch_2(\CE_1) -\half J^2\ch_1(\CE_1) J
\hat \ch_2(\CE_2)\\
& + & J\cdot \hat \ch_2(\CE_2) \hat \ch_3(\CE_1) - J\cdot \hat
\ch_2(\CE_1) \hat \ch_3(\CE_2)
\end{eqnarray}
where we have set $B=0$ for simplicity. The spin is computed from:
\begin{equation}
\langle \Gamma_1, \Gamma_2\rangle = r_1 r_2 \biggl(\mu_1\cdot \hat
\ch_2(\CE_2) - \mu_2 \hat \ch_2(\CE_1) + \frac{\hat
\ch_3(\CE_1)}{r_1} - \frac{\hat \ch_3(\CE_2)}{r_2} \biggr).
\end{equation}

Suppose now that $r,r_1,r_2>0$ and suppose that in some region of
K\"ahler moduli space a sheaf $\CE_1$ of rank $r_1$ destabilizes
$\CE$. That is, we can write
\begin{equation}
0 \rightarrow \CE_1 \rightarrow \CE \rightarrow \CE_2\rightarrow 0.
\end{equation}
Then the standard slope-stability wall is
\begin{equation}
\frac{J^2\ch_1(\CE_1)}{r_1} =\frac{J^2\ch_1(\CE_2)}{r_2}
\end{equation}
and the   walls given by the vanishing of (\ref{D6Wall}) indeed
asymptote to this wall, but again, different distributions of D0D2
charge between the two constituents lead to distinct walls which all
asymptote to a common slope-stability wall.  Therefore our formula
for $\Delta\Omega(x,y)$ gives us some information on the Hodge
polynomials of the cohomology classes that are lost and gained
across this wall.

\subsection{$D4 \to D6 + \overline{D6}$}

One special case of particular interest, which played an important
role in \cite{Denef:2007vg}, occurs when $r= r_1  + r_2=0$.  That
is, the decay of a $D4D2D0$ system into a $D6\overline{D6}$ system.
Here, for $B=0$ we have:
\be
\begin{aligned}
{\rm Im}(Z_1\overline{Z_2})&= \frac{r_2 J^3}{12}  J^2\ch_1(\CE) -
\frac{r_2 J^3}{6} \hat \ch_3(\CE) +\half J^2\ch_1(\CE_2) J \hat
\ch_2(\CE_1)\\
 &- \half J^2\ch_1(\CE_1) J \hat \ch_2(\CE_2)+\cdots\\
\end{aligned}
\ee
Since $\ch_1(\CE)$ is an effective class  (being Poincar\'e dual to
the cycle where the D4 wraps)  the wall cannot extend to infinity,
at least not with $B=0$. However, for large D0 charge $\hat
\ch_3(\CE)$ the wall can be brought to the regime of large $J$ where
our approximations apply.

\section{Discussion}

In this paper we have shown that the physical wall crossing formula
 applies in a more general context than was used in  \cite{Denef:2007vg}.
In particular, combining it with the description of D-branes in
terms of coherent sheaves leads to rather nontrivial agreement with
wall-crossing formulae in the mathematics literature. Moreover, this
discussion suggests some interesting expectations for  a future
theory of the moduli space of stable objects in the derived
category.

One point which should be stressed is the following.  The decays
discussed in section \ref{sec:D4toD4D4} have a potentially important
implication for the OSV conjecture \cite{Ooguri:2004zv}, since one
can arrange that the D4 branes wrap surfaces $S,S_1,S_2$ all of
which are ample, and yet the decay wall is in the K\"ahler cone.
\footnote{Note that in the examples studied in section
\ref{sec:CompareG} $S$ is not ample. Examples where an ample $D4$
decays  into a pair of ample $D4$'s have been independently
discovered in \cite{DenefVandenBleeken}.} This means that
\begin{equation}
\lim_{J\to \infty} \Omega(\Gamma;B+iJ)
\end{equation}
is not well-defined, even for D4D2D0 systems where the D4-brane
wraps and ample divisor!  The wall-crossing formula shows that the
jumps in $\Delta \Omega$ are corrections potentially just as large
as the world-sheet instanton corrections in the refined version of
the OSV conjecture described in \cite{Denef:2007vg}. We defer a
careful examination of this possibility to future work.

 There is one aspect of our discussion
 which is quite unsatisfactory. This becomes apparent upon a more
detailed examination of which states decay as one crosses the wall.
To focus the discussion let us return
 to the case $r=2$ where bundles are
destabilized by exact sequences such as (\ref{E12}). Let us assume
for simplicity that the destabilizing subspaces
$\CV_{1,2}=\CP_{1,2}$ in equations (\ref{eq:projbundleA}) and
(\ref{eq:projbundleB}).

In the mathematical description the change of moduli space is given
by a simultaneous blow-down of $\IC P^{K_{21}-1}$ and blow-up of
$\IC P^{K_{12}-1}$. In other words,  we lose $K_{12}$ states and
gain $K_{21}$ states for a net change of $I_{12} = K_{12}-K_{21}=
\langle \Gamma_1,\Gamma_2\rangle$ states. In the physical
description, on the other hand, a spin $\half (\vert I_{12}\vert
-1)$ multiplet of BPS states moves off to infinity in fieldspace
along a Coulomb branch.

These are very different pictures of what happens to the space of
BPS states as $t$ crosses the wall, although both pictures agree on
the \emph{net} change of BPS states.  Resolving this puzzle is
beyond the scope of the present paper, but we believe the resolution
will be important and might have a significant impact upon our
understanding of the relation between D-branes and constructions in
algebraic geometry.

\acknowledgments

We would like to thank Frederik Denef for collaboration on related
matters and for very useful discussions.  We would also  like to
thank Ron Donagi, Michael Douglas, Lothar G\"ottsche, Juan
Maldacena, Tony Pantev and Richard Thomas for discussions. We owe
special thanks to Lothar G\"ottsche and Richard Thomas for making
their work available to us prior to publication and to Frederik
Denef, Lothar G\"ottsche, and  Richard Thomas for comments on the
draft. This work was partially supported by the DOE under grant
DE-FG02-96ER4094 and the NSF under grant PHY-0555374-2006.

\end{document}